\documentclass[12pt,oneside]{article}

\usepackage{amsfonts,amssymb,graphicx}
\usepackage{amsmath}

\usepackage{mdwlist}
\usepackage{enumitem}

\setlist{nolistsep}

\usepackage[bottom]{footmisc}

\usepackage{caption, subfig}

\usepackage{booktabs}

\def\E{ {\rm E} }

\def\ra{\rightarrow}

\def\no{\noindent}
\def\be{\begin{equation}}
\def\ee{\end{equation}}
\def\bea{\begin{eqnarray}}
\def\eea{\end{eqnarray}}

\def\<{\langle}
\def\<{\langle}
\def\>{\rangle}

\def\ds{\displaystyle}

\def\~{\tilde}

\def\ds{\displaystyle}

\makeatletter
	\def\blfootnote{\xdef\@thefnmark{}\@footnotetext}
\makeatother

\begin{document}
%


\begin{center}

{\bf\sc\Large reproduction and ferromagnetism \\}

\vskip .5truecm
{Ignacio Gallo \\
} 
\end{center}

\blfootnote{email: {\it ignacio@pofume.org}}




\vskip 1 truecm

\begin{abstract}\noindent
\footnotesize I present a stochastic population model that combines cooperative interactions of the type often used
		     in physics with the process of reproduction and death familiar to biology, and I refer to reasons why
		     such interlocking may be of interest to both fields. 
		      \end{abstract}

\vskip 1 truecm

%
%

%
%

%
%

\noindent Physics and biology have developed good frameworks to model the process of stochastic change in systems which experience the action of 
cooperation, and reproduction, respectively. These two frameworks overlap substantially, and it is surprising that neither the physical or the
biological literature seem to consider models interlocking the two effects.

Considering a system that exhibits both reproduction and cooperation is particularly significant in view of the fact that
the conventional explanation of ferromagnetism as a cooperative effect, arising somehow out of quantum mechanical assumptions, has 
repeatedly been called into question \cite{bozorth, feynman, crangle}. Indeed, it has been closely argumented recently, on empirical grounds, 
that the ferromagnetic transition should be interpreted as one due to change through nucleation and growth in the material's crystal structure, with 
cooperative magnetics effects playing no significant role \cite{mnyBook, mnyPaper}.

However, even though the assumptions of quantum mechanics don't offer a good premise for the existence of molecular cooperation either 
from an empirical or a conceptual point of view \cite{aharoni}, it is unsatisfactory to rule out the possibility of the cooperative effect altogether, 
especially in view of recent advances in biology. 

There are obvious analogies linking the growth processes of bacterial colonies and of crystals. On the other hand, bacterial species
have been found to coordinate their activity through several types of cooperative interactions, that have been characterised in mechanistic detail \cite{quorum}.
Since cooperation is the traditional explanation for the ferromagnetic phenomenon, and since crystal nucleation and growth are currently being proposed as
a radically alternative cause, it is sensible to consider that both effects may be involved in physical
materials, in analogy to what is observed in biological systems.

In other words, evidence suggests that both physical and biological processes may involve the intertwined action of reproduction and 
cooperation. In this paper I consider one of simplest ways to model this, which can be described in a nutshell as a superposition of Moran's
random-walk model of genetic drift through death and birth \cite{moran}, and the Glauber dynamics of a mean field ferromagnet \cite{glauber}. 

I will now introduce the
process, and use the subsequent section to characterise the stochastic equilibrium that arises from it. The characterisation gives a full summary
of the possible regimes, and allows some considerations regarding the significance of the model.

\vskip .4 in

\section{The process}

A particularly good scenario for modelling the interlocking of reproduction and cooperation is given by the
possibility, considered recently \cite{rok}, that bacteria living in colonies may be able to regulate the rate at which
their genetic material mutates through the use of {\em quorum sensing}, a cooperative social effect that has 
been characterised in detail in the context of gene expression \cite{quorum}.

We can model this situation in an elementary way by considering a population consisting of two types of bacteria $A$ and $B$ having 
similar survival and reproductive abilities. Types $A$ and $B$ may change into one another through mutations, that may 
come about due to the persistent chemical interactions in which the genetic material is involved. In this context, the existence of DNA 
repairing processes \cite{repair} may render unmutated individuals mechanistically prone to revert mutants to their original form, thus making cooperation
an integral part of the mutation process.

This modelling approach involves very strong assumptions on the nature of the bacterial life-cycles (which are assumed
to be completely uniform), the dependence on the environment (which is neglected), and on population size and 
structure (which are taken as fixed, and trivial, respectively). In addition to this, I'll also assume that rates of mutation are 
the same in both directions $A \ra B$ and $B \ra A$.

Such an elementary setting allows to fully characterise the type of stochastic equilibrium attainable when both
reproduction and cooperation are at work, and can be given both physical and biological interpretations. In physical terms, 
reproduction lends itself to be interpreted in terms of crystal growth, whereas mutations can be seen as configurational changes 
due to thermal noise, such as magnetic spin flips.

\vskip .3 in

\subsection{Unfolding of the process}

\begin{figure}[t]
\centering
	\subfloat[][]{\includegraphics[width=9 cm]{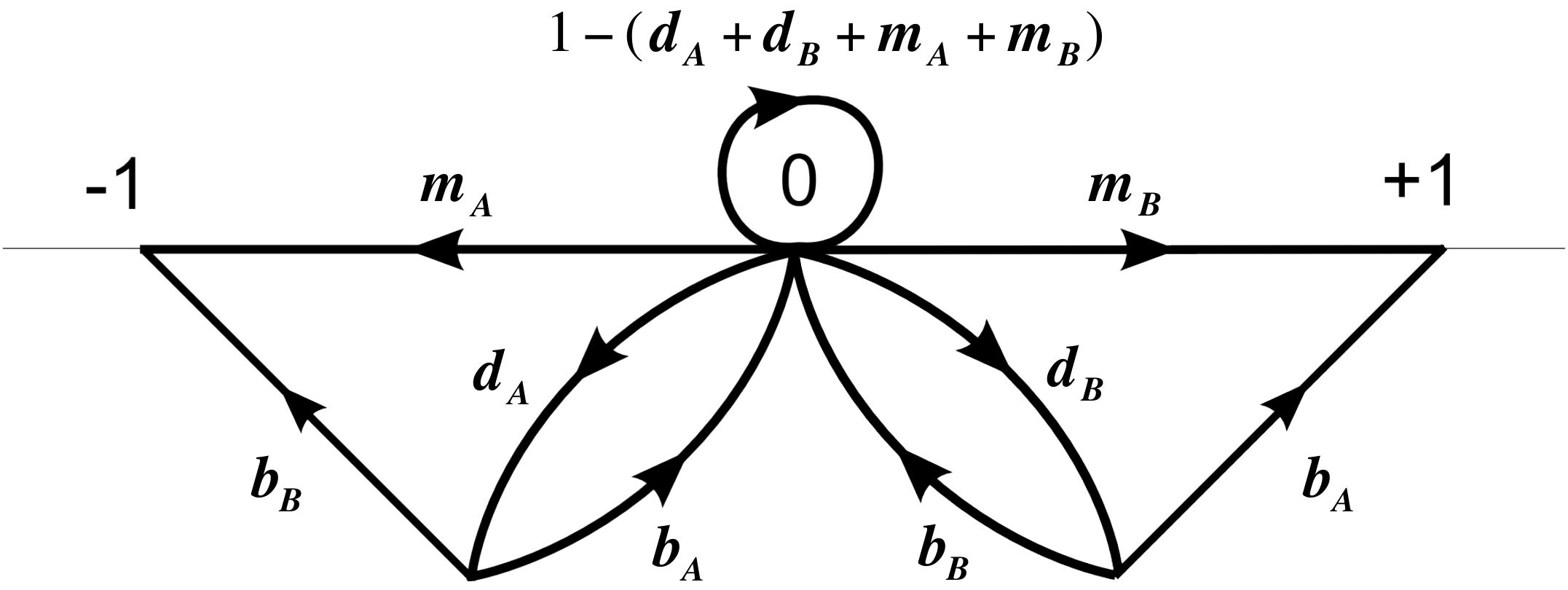}\label{me}}
\\
	\subfloat[][]{\includegraphics[width=7.5 cm]{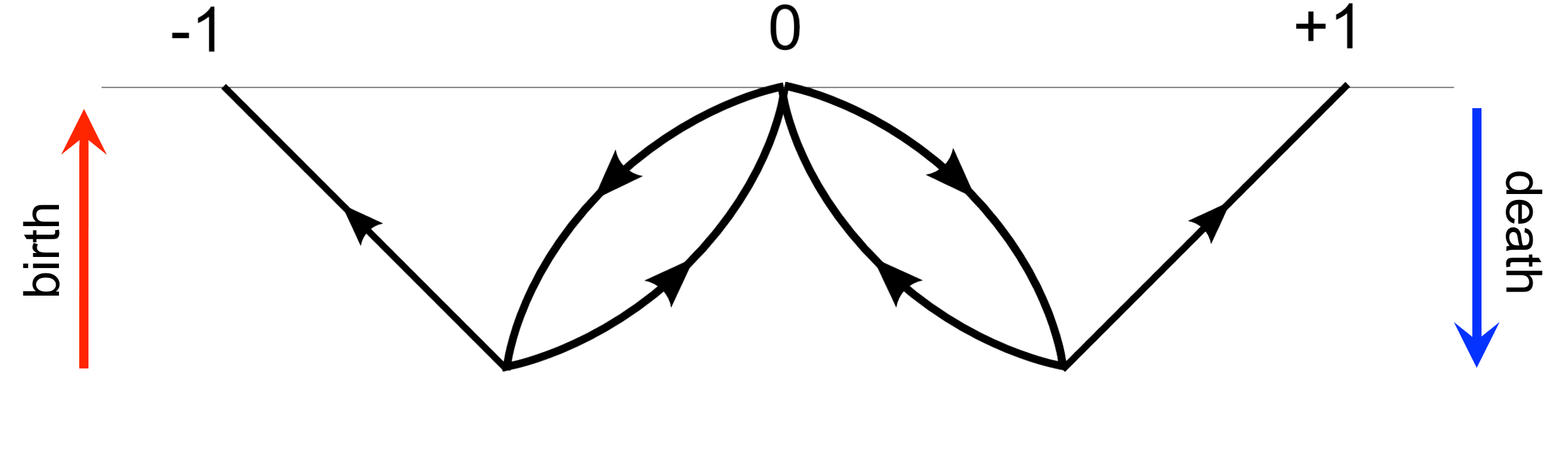}\label{moran}}
	\qquad
	\subfloat[][]{\includegraphics[width=6.8 cm]{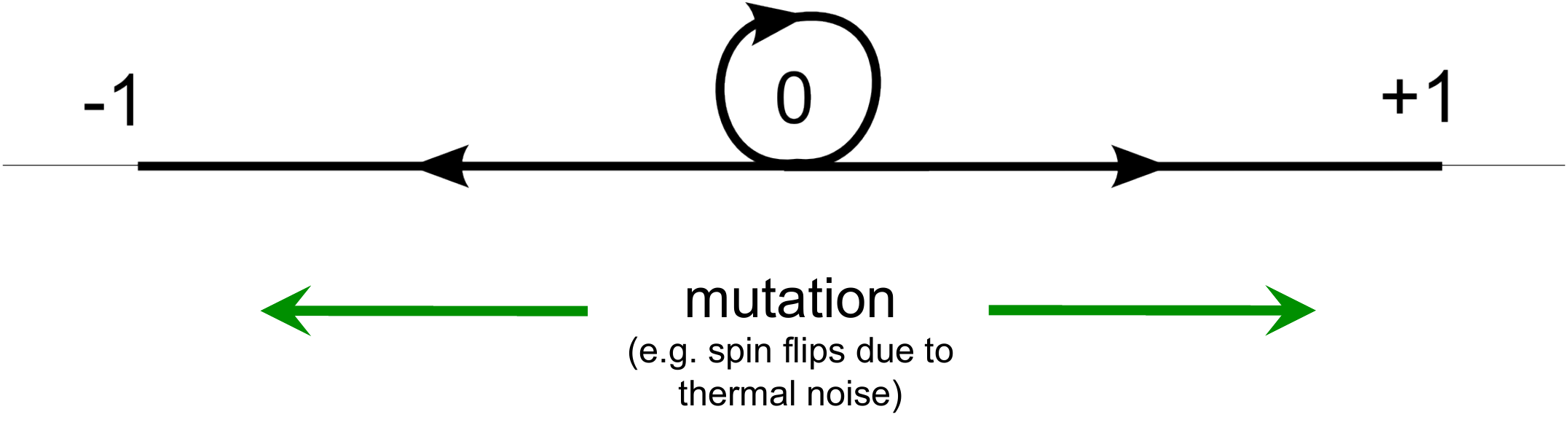}\label{magnet}}
	\caption{\it (a) the random change in the number $X$ of type $A$ bacteria at a given time interval. This process can be see as the 
			superposition of: (b) a Moran random-walk process of genetic drift through death and birth, and (c) the Glauber dynamics of a
			mean-field spin system. }\label{process}%
\end{figure}

The process I want to consider is characterised diagrammatically by Figure \ref{me}, which represents the 
stochastic change in the number $X$ of bacteria of type $A$ present in the population during a given time interval. Since the population 
size is kept fixed at a value $N$, the number of bacteria of type $B$ is given by $N-X$.

Figure \ref{process} shows how the process can be built naturally as a superposition of the Moran process of genetic 
drift (Fig. \ref{moran}), and the Glauber dynamics of a mean-field ferromagnet~(Fig. \ref{magnet}). 

At each time interval an individual is chosen to either live or die. 
If it dies, a new individual is immediately born by one of the survivors to take over the environmental spot 
just vacated. This is an elementary way to model a process of selection deriving from the 
presence of reproduction and death. 

Whereas a birth always systematically follows every death, no deaths might happen during a given time interval: this has been called 
a ``musical chairs" type of process, since it bears analogies with children's playground game \cite{binmore}, and it has been shown that such 
inclusion of time intervals where no deaths happen 
gives a chance to model biological function, in a way that is reminiscent of the modelling of molecular activity in statistical 
thermodynamics \cite{gallo}. 
The biological function that is interesting for the present case is the ability to repair damaged DNA.

When an individual doesn't die, it keeps performing its metabolic processes. These may bring about mutations, and in
particular mutations that may cause a change in the bacterial type, say from $A$ to $B$.
Generally, when a mutation happens in an 
individual, several other unmutated individuals of the type A are left in the population. 
Due to the existence of DNA repairing processes \cite{repair}, 
it's easy to envision ways in which unmutated bacteria of the original type $A$ may be prone to revert a mutant to its original form. This confers 
social dimension to the process of mutation, bringing it close to the standard interpretation of ferromagnetism, though putting the cooperative 
effect on firmer mechanistic grounds.

\vskip .3 in

\subsection{Transition probabilities}

Having established the structure of the process, its precise form is specified by choosing the transition probabilities in Fig. \ref{me}.
I'll define the probabilities as follows, and justify them within the interpretation of the system as a bacterial population.

\vskip .2 in

\begin{tabular}{l|l|l}
{\bf death probabilities} \  & {\bf birth probabilities} \ & {\bf mutation probabilities}  \\
$d_A = x \, d$   & $b_A = x$ & $m_A = (1-d) \, u \, x \, r^{\, N \, x}$  \\
$d_B = (1-x) \, d$   & $b_B = 1-x$ & $m_B = (1-d) \, u \, (1-x) \, r^{\, N \, (1-x)}$  
\\
\end{tabular}

\vskip .2 in

%
%

For example, the probability $d_A$ that an individual of type $A$ dies at a particular time interval is equal to $x \cdot d$. 
This is the probability that an individual of type $A$ is chosen to die ($x=X/N$), multiplied by the probability 
that it actually dies, which I'll call $d$. Parameter $d$ therefore allows to tune the average lifespan of our bacteria, which gives $T=1/d$,
it being the average of a geometric distribution with parameter $(1-d)$.

A more satisfactory way to derive $d_A$ would be to consider that at each time interval any number of bacteria may die due to their 
own individual 
life-cycles, and consequently choose a time-scale that makes multiple simultaneous deaths very unlikely. Such approach is laborious in terms of notation,
though it better shows the relation between the bacterial life-cycles and the population process. The more general interpretation, however, 
reduces to the naive one for our simple case.

Probability $d_B$ can similarly be shown to be equal to $(1-x) \cdot d$. It is also possible to differentiate the lifespans of types
$A$ and $B$ by using different values of $d$ for the two, and I showed in \cite{gallo} that this leads to non-trivial consequences. 
The stress of the present paper is however on
the interplay between reproduction and cooperation, so I'll just assume that all the bacteria are phenotypically equivalent, and
in particular that they share the same life-expectancy $T=1/d$.

Since all bacteria have equivalent phenotypes, and since I'm assuming a ``musical chairs" process where a birth systematically follows each
death, it's easy to see that the birth probabilities are simply $b_A=x$ and $b_B=1-x$.

We finally have the mutation probabilities $m_A$ and $m_B$, which characterise the cooperative nature of the process. For example, the probability
$m_A$ that an individual of type $A$ mutates into one of type $B$ is given by
\begin{equation}\label{mut}
	m_A = (1-d) \, u \, x \, r^{\, N \, x}, 
\end{equation}
where the factors have the following interpretations:

\begin{tabular}{ll}
\\
$(1-d)$   & probability that no death happens (so a mutation might arise)  \\
$u$   & probability of a mutation  which leads to a change in bacterial type,  \\
$x$   & probability that the mutant's original type happens to be $A$,   \\
$ r^{\, N \, x}$   & probability that the mutation is not reverted to the original type $A$.
\\
\\
\end{tabular}

\noindent The factor $ r^{\, N \, x}$ is the typical transition probability for the Glauber dynamics of a mean-field 
ferromagnet \cite{vankampen}.
However, whereas within statistical physics such factor is usually justified as the simplest choice which is consistent with
the Gibbs distribution \cite{glauber}, the process of DNA repair provides a direct mechanistic justification for our choice of stochastic 
dynamics.

We may in fact assume that an interaction between a mutant and a non-mutant may revert the mutant to its original type with
probability $(1-r)$. Assuming the mutant's interactions with non-mutants to be independent events, this gives a probability
$
r^X
$
that all interactions fail to revert the mutant. Therefore, keeping in mind that the total number of type $A$ bacteria is equal to $X=N \, x$,
we obtain (\ref{mut}) for the probability that a mutation from $A$ to $B$ happens and is not repaired.

If we now define the stochastic change in $X$ at a given interval $t$ as
$$
\Delta X = X_{t+1} - X_{t},
$$
we can use inspection on Fig. \ref{me} to find the first two moments of $\Delta X$,
\begin{eqnarray*}
	\E \big [  \Delta X \big ] &=&   ( m_B + d_B \, b_A ) - ( m_A + d_A \, b_B ) =
\\
					    &=&	(1-d) \, u \big [ (1-x) \, r^{N(1-x)} - x \, r^{N x} \big  ],
\\	
	\E \big [   ( \Delta X ) ^2 \big ] &=& ( m_B + d_B \, b_A ) + ( m_A + d_A \, b_B ) =
\\	
	&=& d \, 2 \, x \, (1-x) + (1-d) \, u \big [ (1-x) \, r^{N(1-x)} + x \, r^{N x} \big  ].
\end{eqnarray*}

Using these we can write down the moments of the stochastic change in the relative frequency $x=X/N$,
\begin{eqnarray*}
	M &=& \E \bigg [  \frac{\Delta X}{N} \bigg ] = \frac{1}{N} \, \E \big [  \Delta X \big ] ,
\\	
	V &=& \E \bigg [ \bigg ( \frac{\Delta X}{N} \bigg)^{\!\!2} \, \bigg ] =  \frac{1}{N^2} \, \E \big [  ( \Delta X ) ^2  \big ].
\end{eqnarray*}
In the next section we'll use these two quantities to characterise the stochastic equilibrium attained by the process.

\section{Characterization of the stochastic equilibrium}

Having obtained $M$ and $V$ in the last section, we now can write down the equilibrium distribution $\phi$ for $\ds x = X/N$ as
\begin{equation}\label{wright}
\phi(x) = \frac{C}{V} \, \exp \bigg \{ 2\int\frac{M}{ V } dx \bigg \},
\end{equation}
where $C$ is the normalisation constant that makes $\phi$ a probability distribution over the interval $0 \leqslant x \leqslant 1$.
This formula was derived for the Wright population genetical process \cite{wright}, where a whole biological population
is regenerated at discrete time intervals (usually called {\em generations}), and it has been shown to apply also to Moran's random-walk 
equivalent of the former \cite{moran}\cite{cannings}. The model I'm considering can be seen as a variation on 
Moran's, which, due to its single-step nature, lends itself to be modified in an intuitive way.

The process I'm considering contains typical features of both biological and physical models: it is in fact useful to apply rescalings
of the parameters typical of both disciplines, in order to identify the most interesting regimes attained by (\ref{wright}).

It is common in population genetics to define a parameter $\theta = N \, u$: this corresponds to the fact that the rate of
mutation of a genetic site $u$ is typically very small compared to the lifespan of an individual organism. Mathematically, this
rescaling prevents the law of large numbers from turning $x=X/N$ into a deterministic variable for large $N$.

Since the death rate $d$ appears explicitly in present process I'll define parameter $\hat \theta$ as
$$
\hat \theta =N \, \frac{(1-d)}{d} \, u.
$$

We'll also make the typical assumption of mean field ferromagnetism, which in this case amounts to defining
$$
\rho = r^N.
$$
Mathematically, we are assuming that $r$ is sufficiently close to $1$ for $r^N$ to stay away from zero at a relevant large $N$,
and this prevents the probability of mutation from vanishing, which would reduce the process to the one shown in Fig. 1(b).

The meaning of $\rho = r^N$ is quite straightforward within the narrative of population biology: in a population consisting of
only one type of organism, $\rho$ gives the probability that there doesn't happen to be any repair event following a mutation event,
and therefore it characterises the cooperative aspect of the process of mutation.

We can now apply our rescalings to $M$ and $V$: notice in particular that the second term of $V$ is of order $1/N$, so it can
be safely neglected in our large $N$ expression of $\phi$, which gives
\begin{equation}\label{distrib}
\phi = \frac{C}{x (1-x)} \exp \bigg \{ \hat \theta \!\! \int\frac{ (1-x) \, \rho^{1-x} - x \, \rho^{x} }{ x \, (1-x) } dx \bigg \}.
\end{equation}

Eq. (\ref{distrib}) can be expressed in the closed form
$$
\phi = C \cdot \frac{ \exp \Big \{ \hat \theta \, \rho \, \Big ( {\rm Ei} ( - x \ln \rho) + {\rm Ei}( - (1-x) \ln \rho ) \Big ) \Big \}  }{x \, (1-x)},
$$
by using the ``exponential integral" ${\rm Ei}(x)$, a special function defined as
$$
{\rm Ei}(x) = \int_{-\infty}^x \frac{ e^t }{t} \, dt.
$$

\begin{figure}[t]
		    \centering
		    \includegraphics[width=15 cm]{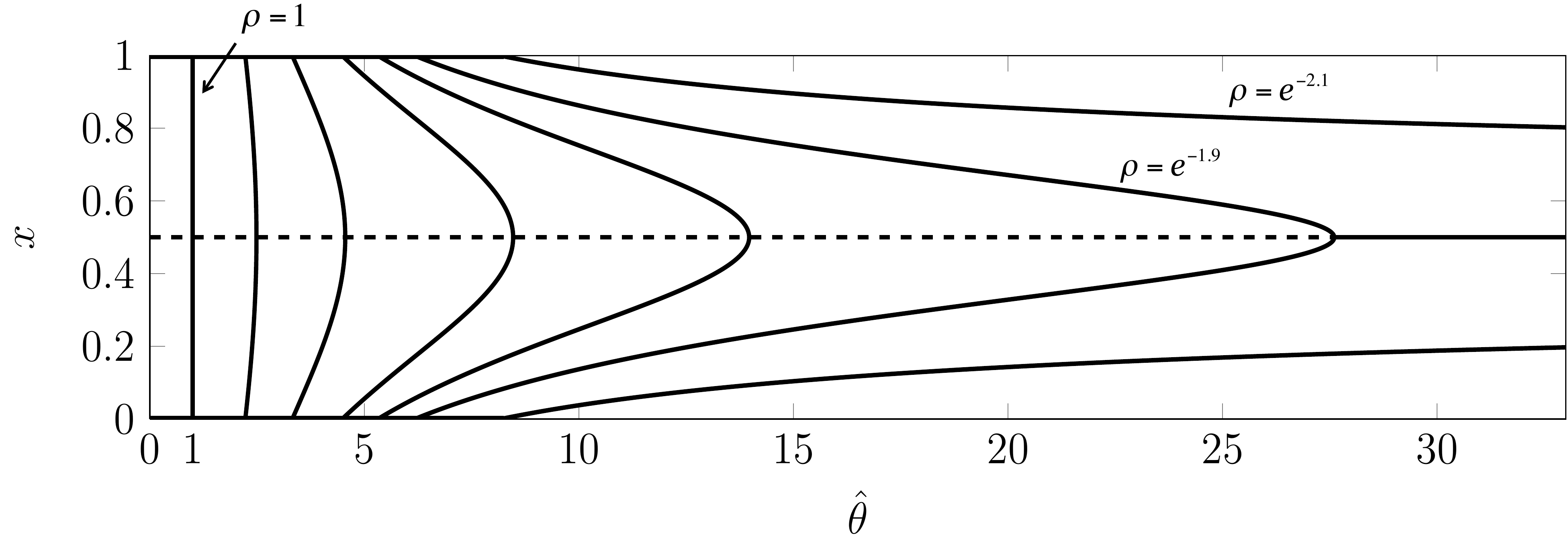}
	\caption{\it Mode diagram: values of $x$ maximising the equilibrium distribution $\phi$, as function of $\hat \theta$, for different values 
			of $\rho$. 
			The diagram shows the transition from a minimum (dashed line) to a maximum (solid line) of the stationary point at $x=0.5$ 
			only for the case $\rho=e^{ \, -1.9}$, which corresponds to Fig. 3. }\label{modeDiagram}		    
\end{figure}

\begin{figure}[t]
	\centering
	\subfloat[][]{
			\includegraphics[height=6 cm]{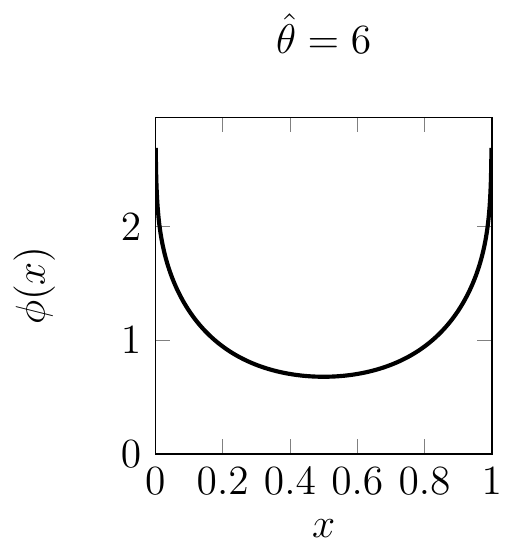}
			}
	\subfloat[][]{
			\includegraphics[height=6 cm]{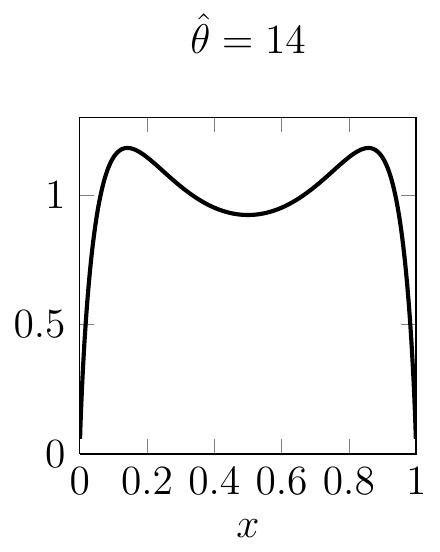}
			}
	\subfloat[][]{
			\includegraphics[height=6 cm]{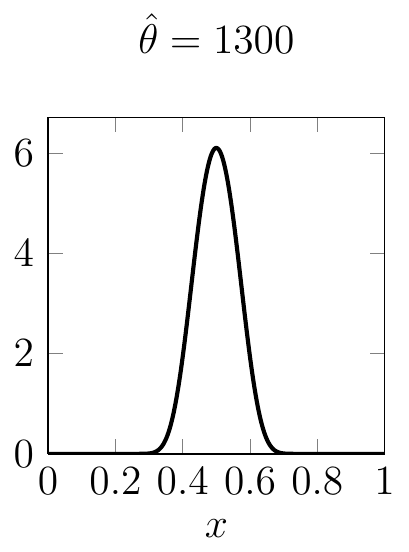}
			}
	\caption{\it The equilibrium distribution $\phi(x)$ at increasing values of the mutation parameter $\hat \theta$ for $\boldsymbol{\rho =  e^{-1.9}}$. 
				The modes at $0$ and $1$, which are typical of low mutation regimes
				in population genetics (case a) ), shift continuously towards $x=0.5$ and the distribution becomes unimodal (case c) ) at 
				a finite value of $\hat \theta$. This behaviour is typical for $\rho <  e^{-2}$.}\label{lowInter}%
\vskip .1 in
	\subfloat[][]{
			\includegraphics[height=6 cm]{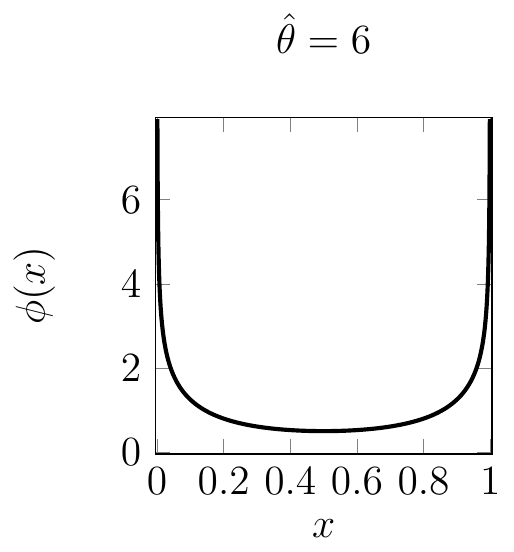}
			}
	\subfloat[][]{
			\includegraphics[height=6 cm]{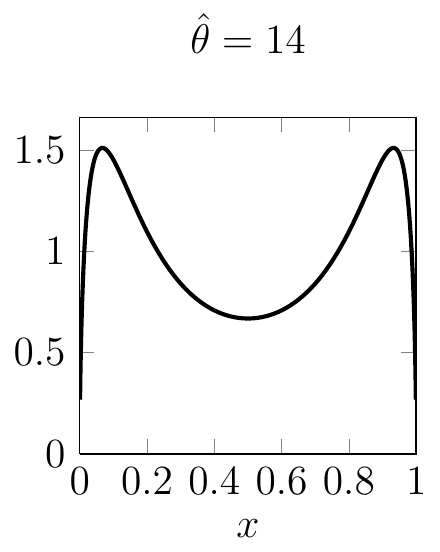}
			}
	\subfloat[][]{
			\includegraphics[height=6 cm]{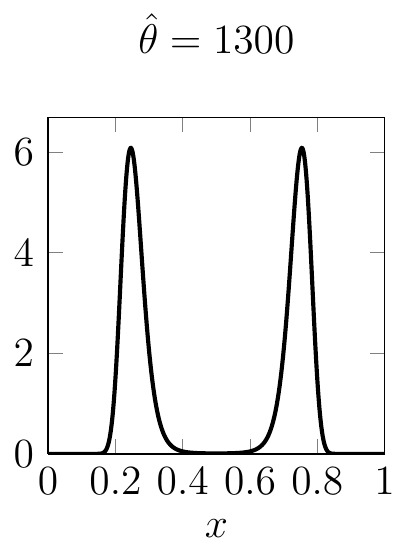}
			}
	\caption{\it $\phi(x)$ at increasing values of $\hat \theta$ for $\boldsymbol{\rho =  e^{-2.1}}$. As in Fig. \ref{lowInter} the modes shift continuously towards $x=0.5$ as $\hat \theta$ increases.
				 However, since $\rho >  e^{-2}$, the distribution remains bimodal for all $\hat \theta$, and admit two values in the limit. All distributions in Fig. \ref{lowInter} and
				  \ref{highInter}  correspond to large $N$ limits, showing that in the chosen scaling the process retains its stochasticity even in large systems.}\label{highInter}%
\end{figure}

In order to categorise the possible behaviours of $\phi$ we can use formula (\ref{distrib}) to get the equilibrium distribution's stationary
points by setting 
$$
\frac{d \phi}{d x } = 0.
$$

Therefore, since
$$
\frac{d \phi}{d x } = \frac{ C}{ V^2 } \cdot  \exp \bigg \{ 2\int\frac{M}{ V } dx \bigg \}   \cdot ( 2 M - \frac{d \, V}{d x }),
$$
stationary points satisfy
\begin{equation}\label{preModeEq}
	2M - \frac{d \, V}{d x } = 0.
\end{equation}

Dividing (\ref{preModeEq}) $2 d$, and keeping in mind that $\hat \theta = N \, u \, (1-d) / d$ and $ \rho = r^N $, the last equation becomes:
$$
\hat \theta \big (  (1-x) \rho ^{1-x} - x \rho ^ x ) - (1-2x) = 0
$$

A convenient way to analyse this equation is to solve for $\hat \theta$ as a function of $x$, and see graphically how this function changes as a
function of the interaction parameter $\rho$:
\begin{equation}\label{modeEq}
	\hat \theta = \frac{1-2x}{(1-x) \rho ^{1-x} - x \rho ^ x }.
\end{equation}

So equation (\ref{modeEq}) can be used to construct the ``mode diagram" in Fig. 2: points in this diagram correspond to maxima of the equilibrium $\phi$. 
These points, however, are {\em modes} rather than thermodynamic {\em phases} since, due to our chosen scalings, the equilibrium $\phi$ remains stochastic in the 
limit $N \ra \infty$. This is demonstrated in Figures \ref{lowInter} and \ref{highInter}, that show the shape of $\phi$ at increasing values of the mutation 
parameter $\hat \theta$ for $\rho =  e^{-1.9}$, and $\rho = e^{-2.1}$, respectively. 

\vskip .5 in

The fact that the distribution doesn't attain a deterministic limit for large $N$ is brought about by the presence of reproduction: in its absence a deterministic limit
is attained, independently of the choice of scaling for the parameters. In practical terms, the absence of reproduction and death eliminates the term $2 \, d \, x(1-x)$ from
$V$, making it impossible for the ratio $M/V$ to remain bounded for large $N$. In particular, the retainment of stochasticity in the thermodynamic limit shows that 
the Gibbsian approach to the study of large systems is not applicable to the present case.

The system however retains the main characteristic deriving from cooperation: Figures~\ref{lowInter} and \ref{highInter} demonstrate how for $\rho < e^{-2}$ -which
corresponds to a high probability of DNA repair events, due to strong bacterial interactions- two 
coexisting deterministic phases are attained in the limit $\hat \theta \ra \infty$, whereas only one is attained for $e^{-2} < \rho \leqslant 1 $.
This fact can be easily derived analytically from Eq. (\ref{modeEq}), and it is also visible in Fig.~\ref{modeDiagram}, from how the curve corresponding to  $\rho = e^{-2.1}$ 
diverges to two horizontal asymptotes.

We have that, in fact, $\phi$ admits bimodal regimes for any value $\rho<1$. However for $\rho> e^{-2}$ the bimodality is lost at a finite value of $\hat \theta$, as
demonstrated in Fig.~\ref{modeDiagram} for the case~$\rho = e^{-1.9}$.

We therefore have that the interlocking of cooperation and reproduction gives rise to an interesting kind of behaviour, 
that may remain stochastic even for systems of large size,
and that shows a robust type of transition from low-mutation 
to high-mutation regimes through an intermediary state (e.g. Fig.~\ref{lowInter}(b)), rather than the abrupt shift exhibited by typical models of population 
genetics~\cite{crowkimura}.

The robustness of the resulting model offers a good testing ground for the idea suggested by the empirical evidence 
mentioned in the introduction, according to which the interplay between reproduction and cooperation may be a characterising feature of processes at
work in a wide range of natural domains.

\vskip .5 in

\noindent {\bf Acknowledgments}  \
This research was supported by a Marie Curie Intra European Fellowship within the 7th European Community Framework 
Programme. Numerical calculations were performed using the free software Maxima.

\end{document}